\documentclass[a4paper,fleqn,usenatbib]{mnras}

\usepackage{mathptmx}

\usepackage[T1]{fontenc}
\usepackage{ae,aecompl}

\usepackage{graphicx}	
\usepackage{amsmath}	
\usepackage{amssymb}	
\usepackage{color}
\usepackage{multirow}

\def\aj{AJ}%
\def\apj{ApJ}%
\def\apjl{ApJ}%
\def\apjs{ApJS}%
\def\ao{Appl.~Opt.}%
\def\aap{A\&A}%
\def\icarus{Icarus}%
\def\mnras{MNRAS}%
\title{The   interaction    of  hydrodynamic    shocks   with
  self-gravitating  clouds.}

\author[S.A.E.G. Falle et al.]{
S. A. E. G.  Falle,$^{1}$\thanks{E-mail: sam@amsta.leeds.ac.uk}
B. Vaidya$^{2}$
T. W. Hartquist$^{3}$
\\
$^{1}$Department of Applied Mathematics, University of Leeds, Leeds
LS2 9JT, UK\\
$^{2}$Dipartimento  di  Fisicia,  University  of  Torino,  Via  Pietro
Giuria, 1, 10125, Torino, Italy.\\
$^{3}$School of  Physics  and  Astronomy,  University of Leeds, Leeds
LS2 9JT, UK
}
\date{Accepted XXX. Received YYY; in original form ZZZ}

\pubyear{2016}

\begin{document}
\label{firstpage}
\pagerange{\pageref{firstpage}--\pageref{lastpage}}
\maketitle

\begin{abstract}
We  describe the  results  of  3D simulations  of  the interaction  of
hydrodynamic  shocks  with  Bonnor-Ebert  spheres  performed  with  an
Adaptive Mesh  Refinement code.   The calculations are  isothermal and
the clouds are embedded in a medium in which the sound speed is either
four or ten times that in the  cloud.  The strengths of the shocks are
such that they induce gravitational collapse  in some cases and not in
others and we derive a simple estimate for the shock strength required
for this to occur.  These results  are relevant to dense cores and Bok
globules  in star  forming  regions subjected  to  shocks produced  by
stellar feedback.
\end{abstract}


\section{Introduction}

It  has long  been  recognised that  the  triggering of  gravitational
collapse by shocks could be  important in star formation. For example,
\cite{Elmegreen:1977p14980} showed  that the  dense layer  between the
shock  and an  ionisation front  can be  gravitationally unstable  and
\cite{Cameron:1977p14977} suggested that  a supernova-induced birth of
the  solar  system  could  account for  the  presence  of  short-lived
radioisotopes  (SLRI)  in  meteorites.   There  is  also  considerable
observational  evidence for  star formation  triggered by  supernovae,
ionisation  fronts,  protostellar outflows  and  other  shocks in  the
interstellar   medium   (e.g.     \citealt{Preibisch,   Lee,   Snider,
  Yokogawa}).
  
There  have  been  a  considerable number  of  simulations  of  shocks
interacting   with    gravitationally   bound   clouds    (see   e.g.
  \citealt{Boss:1995p12081,    Foster:1996p11242,   Foster:1997p14940,
    Boss:1998p14945,      Vanhala:1998p11237,      Vanhala:2002p14946,
    Boss:2008p14960,         Leao:2009p13764,         Boss:2010p12242,
    Boss:2010p12246,     Boss:2012p12244,     Gritschneder:2012p13767,
    Boss:2013p13775,  Li:2014}).  These  do, indeed,  show that  shock
  waves can trigger  gravitational collapse, as long  as the radiative
  cooling time is sufficiently short.

In  a  previous  paper  \cite{Vaidya:2013p12979},  we  considered  the
interaction  of isothermal,  plane parallel  shocks with  magnetically
sub-critical clouds.  Although gravitational  collapse cannot occur in
such  clouds in  the absence  of ambi-polar  diffusion, we  found that
shocks with  Alfv\`enic Mach  numbers of 2  could produce  a temporary
increase in  the density  by a  factor of  $10^3$.  This  is due  to a
combination   of  shock   focusing   and  MHD   effects  rather   than
gravitational collapse.

In this paper we will consider the purely hydrodynamic version of this
problem, that  of an isothermal  shock interacting with  an isothermal
Bonnor-Ebert sphere.  This is exactly  the same as that  considered by
\cite{Li:2014},  but they  only considered  a small  number of  cases,
whereas our  purpose is to  derive a  simple expression for  the shock
strength required to induce  gravitational collapse. The astrophysical
application   that  we   have  in   mind   is  that   of  the   dense,
quasi-stationary cores  found in  star-forming regions, which  in many
cases   appear   to   be   close   to   Bonnor-Ebert   spheres   (e.g.
\citealt{Schnee}).  These  are presumably gravitationally  stable, but
their collapse could  be triggered by shocks due to  stellar winds and
jets, ionisation fronts and supernovae.

We  will  also   briefly  consider  the  effect   of  self-gravity  on
Kelvin-Helmholtz and  Richtmyer-Meshkov instabilities since  these may
be important for  cloud destruction and mixing of SLRIs  into gas that
will       form        a       protoplanetary        system       (see
e.g. \citealt{Boss:2013p13775}).

Section \ref{numtech}  describes the numerical method  and initial and
boundary conditions.  The general evolution of the clouds is discussed
in  section  \ref{shocks}  and  a comparison  of  the  development  of
non-gravitational   instabilities   in   models   with   and   without
self-gravity is presented in section \ref{instab}. Section 5 concludes
the paper.

\section{Numerical Method and Initial Conditions}
\label{numtech}
\subsection{Numerical code}
The calculations  were performed  with the hierarchical  adaptive mesh
refinement  (AMR) code MG  \citep{Falle:2012p12513}.  This  solves the
equations  of  hydrodynamics  using   a  second  order  upwind  scheme
described  in \citep{Falle}.  A  hierarchy of  $n$ grids  levels, $G_0
\cdots G_{n-1}$,  is used, and the  mesh spacing for  $G_n$ is $\Delta
x/2^{n}$, where  $\Delta x$ is the  cell size for  the coarsest level,
$G_0$.  $G_0$ and $G_1$ cover  the entire domain, but finer grids need
not do so.  Refinement is on a cell-by-cell basis and is controlled by
error estimates based on the difference between solutions on different
grids, i.  e.   the difference between the solutions  on $G_{n-1}$ and
$G_n$  determine refinement  to $G_{n+1}$.   Self-gravity  is computed
using a full approximation multigrid to solve the Poisson equation.

\begin{figure*}
\includegraphics*[viewport=16 79 397 541,
width=2.1\columnwidth]{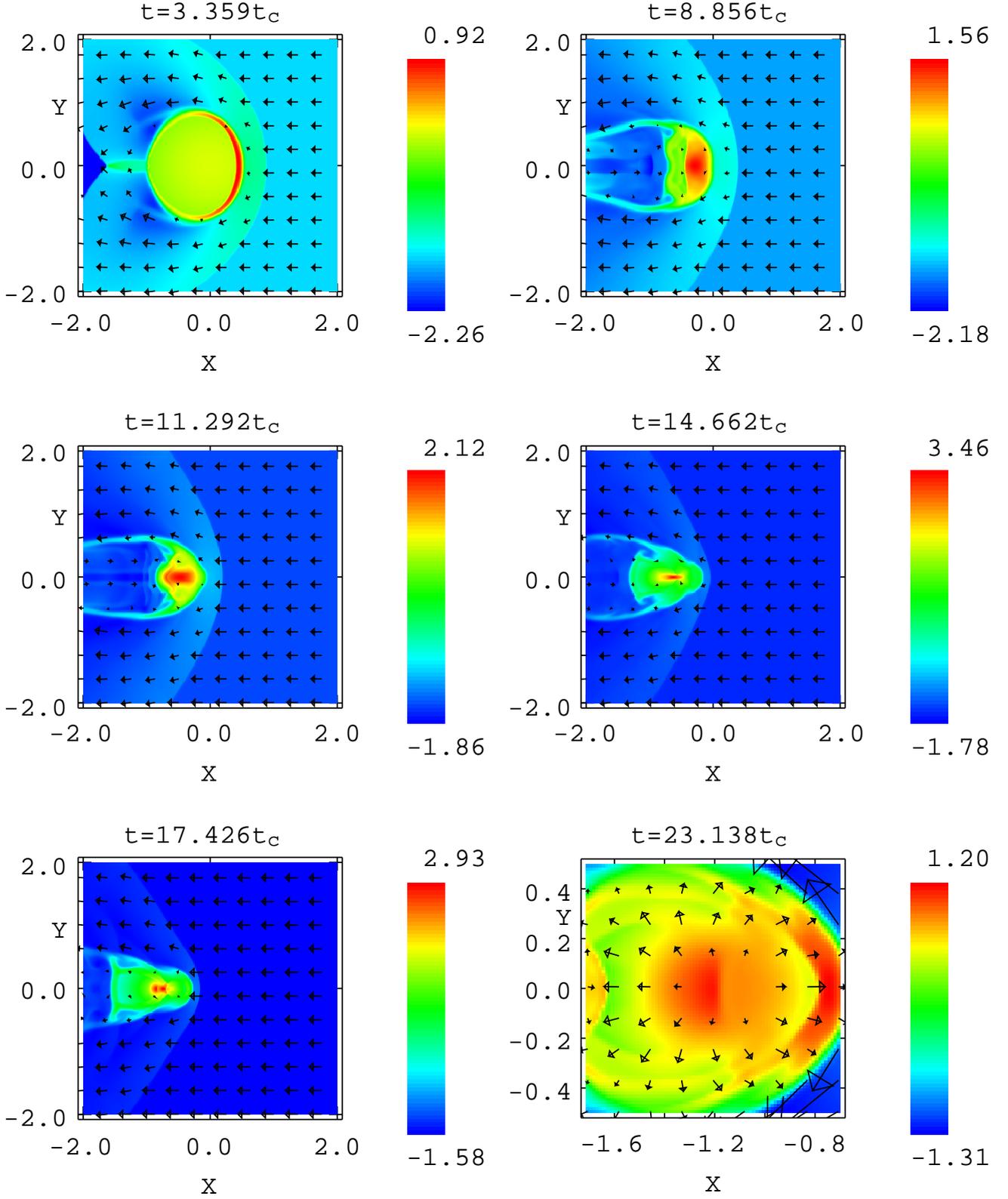}
\caption{Logarithm of the ratio of  the density to the central density
  in the equilibrium state and velocity  arrows in the z = 0 plane for
  the  State 1, $M_s  = 2.2$  simulation. The  velocity arrows  in the
  bottom right panel are in the rest frame of the dense region.}
\label{fig1}
\end{figure*}

\begin{figure*}
\includegraphics*[viewport=16 79 397 541,
width=2.1\columnwidth]{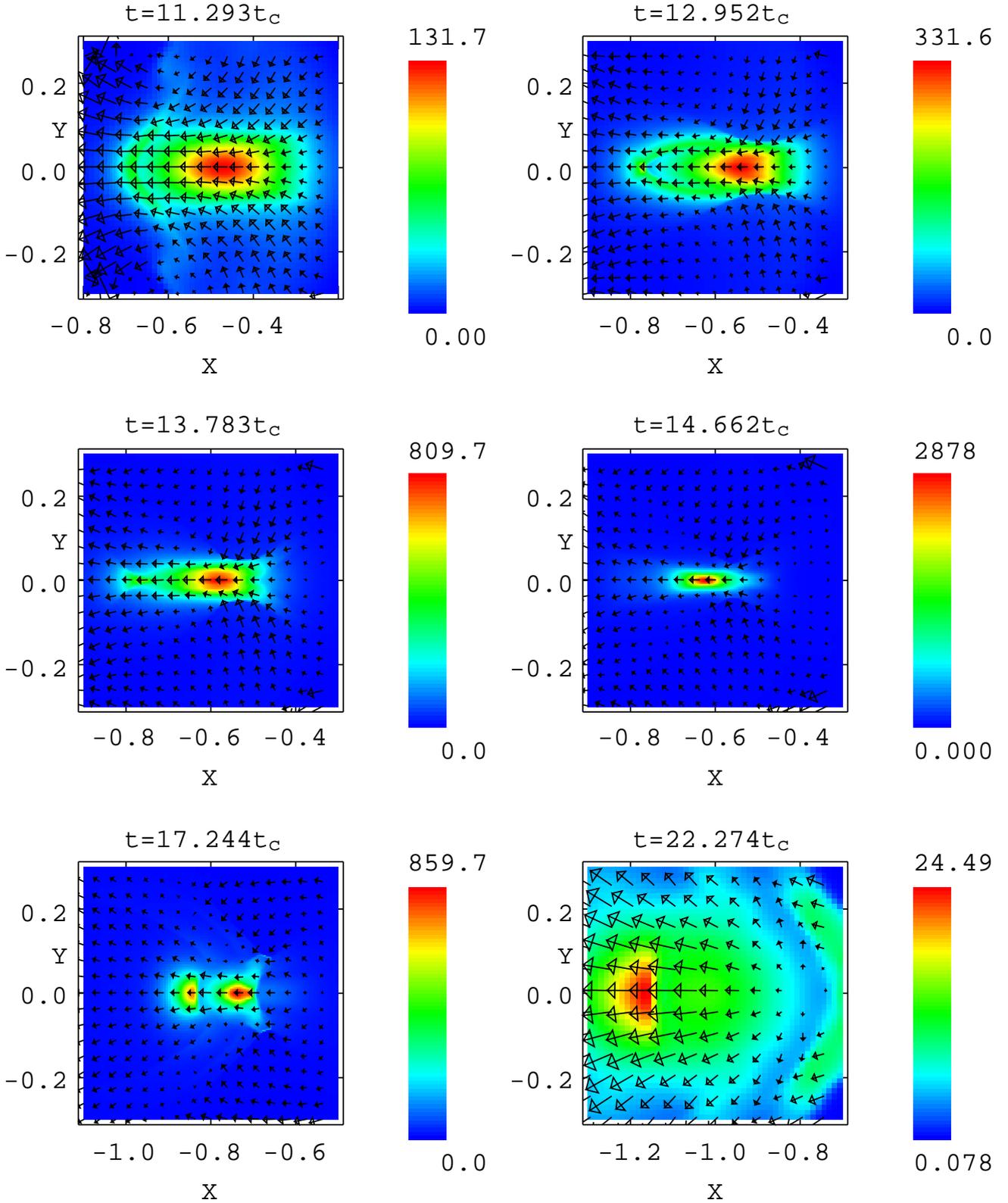}
\caption{Detail of the shock convergence for  the State 1, $M_s = 2.2$
  simulation. Same as Fig.~\ref{fig1} except that the density scale is
  linear and all  the velocity arrows are in the  initial rest frame of
  the cloud}
\label{fig2}
\end{figure*}

\begin{figure*}
\includegraphics*[viewport=16 79 397 541,
width=2.1\columnwidth]{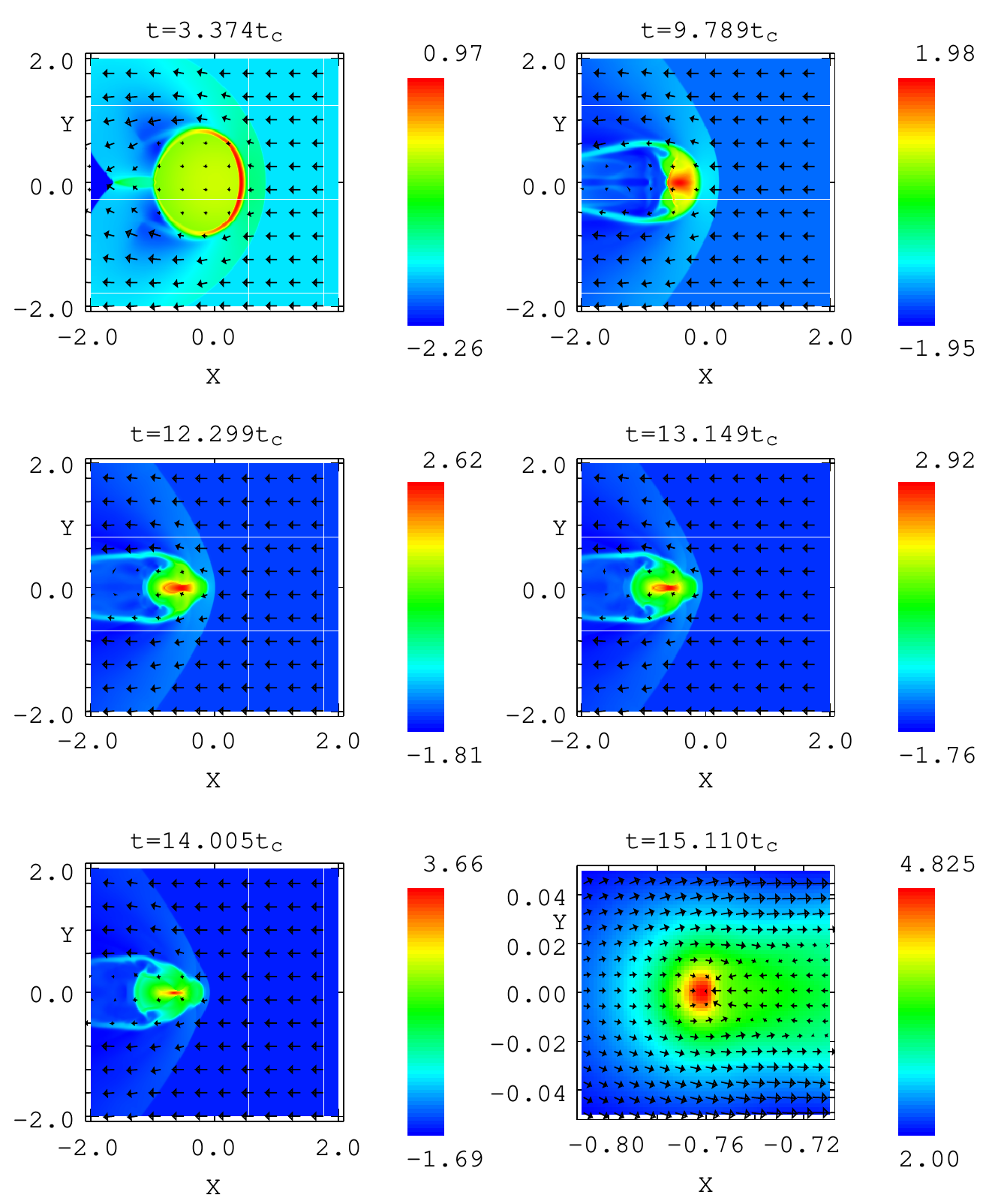}
\caption{Logarithm of the ratio of  the density to the central density
  in the equilibrium state and velocity  arrows in the z = 0 plane for
  the  State 1, $M_s  = 2.3$  simulation. The  velocity arrows  in the
  bottom right panel are in the rest frame of the dense region.}
\label{fig3}
\end{figure*}

\subsection{Domain and Grids}

\begin{table*}
\centering
\begin{tabular}{|c|c|c|c|c|c|c|c|c|c|}
\hline State  & $\rho_c$ &  $P_e$ &  $c_e$ & $M_1$  & $M_2$ &  $M_s$ &
$\rho_{max}/\rho_c$ & Maximum Resolution & Behaviour\\
\hline\hline
1 & 0.37  & 0.2 & 10.0 &  2.65 & 1.91 & 2.1 & $1.74~10^3$  & $1280^3$ &
Rebound \\
 & & & & & & 2.2 & $2.88~10^3$ &  $1280^3$ & Rebound \\
 & & & & & & 2.3 & - &  $2560^3$  & Collapse \\
\hline
2 & 0.37  & 0.2 & 4.0 & 2.65  & 1.91 & 2.0 & $2.88~10^2$  & $1280^3$ &
Rebound \\
& & & & & & 2.1 & $5.55~10^2$ &  $1280^3$ & Rebound \\
& & & & & & 2.2 & - & $1280^3$  & Collapse \\
\hline
3 &  1.09 &  0.45 & 4.0  & 1.76  & 1.66  & 1.6 &  $91.6$ &  $1280^3$ &
Rebound \\
 & & & & & & 1.7 & - & $1280^3$ & Collapse \\
\hline
4 & 2.16 & 0.7 & 4.0 & 1.41 & & 1.4 & $72.4$ & $1280^3$ & Rebound
\\
 & & & & & & 1.5 & - & $1280^3$ & Collapse \\
\hline
5 & 3.9 & 0.95 & 4.0 & 1.21 & &1.2 & $ 20.6$ & $1280^3$ & Rebound \\
 & & & & & & 1.3 & - &  $1280^3$ & Collapse \\
\hline\hline
\end{tabular}
\caption{Simulation parameters.  Here $\rho_c$  is the central density
  in the equilibrium Bonnor-Ebert  sphere, $\rho_{max}$ is the maximum
  density attained, $M_s$ is the  incident shock Mach number, $M_1$ is
  the shock Mach number for which  the post-shock pressure is equal to
  $1.4$ and $M_2$ is the Mach  number for which the pressure behind the
  bow shock is equal to $1.4$.}
\label{table1}
\end{table*}

Although the problem is  axisymmetric, the calculations were performed
on a three-dimensional  Cartesian grid.  This saves us  the trouble of
writing  an axisymmetric  Poisson solver  and  has the  merit that  it
allows for non-axisymmetric instabilities that are sufficiently strong
to be triggered by rounding error. The domain is $-2 \le x \le 2$, $-2
\le y \le 2$, $-2 \le z \le 2$, with the centre of the cloud initially
at the  origin.  Initially $6$  grids were  used with a  resolution of
$10^3$  on  $G_0$, which  gives  an  effective maximum  resolution  of
$320^3$. Note  that $G_0$ needs to  be coarse in order  to ensure fast
convergence of the multigrid Poisson  solver.  This resolution is more
than  adequate for  the equilibrium  state, but  is not  sufficient to
resolve  the  high  density  region  that is  produced  by  the  shock
interaction.   However,  the code  has  the  ability to  resolve  such
regions  by increasing  the  number of  refinement  levels during  the
course of  the calculation (see  Table \ref{table1} for  the effective
resolution in each run).

\cite{Truelove}  have  pointed  out that,  in  calculations  involving
gravitational collapse, one needs to resolve the Jeans length

\begin{equation}
\lambda_J = \left(\frac{\pi c^2}{G \rho} \right)^{1/2},
\label{jeans}
\end{equation}

\noindent
where $c$ is  the sound speed and $\rho$ is  the density. They suggest
that the  mesh spacing needs  to be  $\simeq 0.25 \lambda_J$  to avoid
artificial fragmentation  when a dense region  moves from a fine  to a
coarse  grid.   This   is  not  a  situation  that   occurs  in  these
calculations,  but  it is  nevertheless  useful  to compare  the  mesh
spacing with $\lambda_J$.

The  highest  density  and  hence  the  smallest  Jeans  length  in  a
non-collapsing case occurs in the State  1 $M_s = 2.2$ calculation and
this gives  $\lambda_J =  0.033$.  The  mesh spacing  in this  case is
$\Delta x =  0.0031$, so the Jeans length is  comfortably resolved. It
is even  better resolved  in the other  non-collapsing cases.   In the
collapsing cases  the calculation is  stopped before the  Jeans length
becomes unresolved,  by which time  it is  clear that the  collapse is
unstoppable.

\subsection{Initial Conditions}

We first  compute the collapse of an  initially uniform, non-rotating,
isothermal, spherical cloud to a stable hydrostatic equilibrium state.
The cloud has a sound speed  $c_c$ and is embedded in a warmer uniform
medium with sound speed $c_e$  and pressure $P_e$.  The cloud material
is tracked with an advected scalar $\alpha$ that is unity in the cloud
and zero in the surroundings.  The sound speed, $c$, is given by

\begin{equation}
c^2 =  \alpha c_c^2  + (1  - \alpha)  c_e^2.
\label{sound}
\end{equation}

\noindent
The scalar is also used to turn off gravity in the external medium.

We use units in which $c_c =  1$, the gravitation constant $G = 1$ and
the mass of the  cloud $m = 1$.  In these  units, the maximum external
pressure  that can  be supported  by a  stable Bonner-Ebert  sphere is
$P_{max} \simeq 1.4$ \cite{Bonnor:1956p12177}. We consider four stable
states with  external pressures  $P_{ext} =  0.95, ~0.7,  ~0.45, ~0.2$
(see  Table~\ref{table1}).  These  states were  generated by  starting
with a uniform  density cloud with $\rho = 1$.   The initial radius is
then fixed by the requirement that the mass of the cloud is unity.

This  initial state  was then  allowed  to collapse  until it  reached
equilibrium.  As  noted by  \cite{Boss:2010p12242}, a cloud  formed in
this way oscillates about the  Bonnor-Ebert state for some time.  This
was prevented by imposing a drag force of the form

\begin{equation}
{\bf F}_d = -A \rho {\bf v},
\label{drag}
\end{equation}

\noindent
where $A$  is a  suitable coefficient and  ${\bf v}$ is  the velocity.
The drag force is switched off once the cloud has become static.

\section{Shock Interaction}
\label{shocks}

Once an  equilibrium Bonner-Ebert sphere  has been obtained,  a planar
isothermal shock  with Mach  number $M_s$ moving  in the  negative $x$
direction is introduced near the cloud. The density and velocity at $x
=  2$ are  fixed  at the  values given  by  the Rankine-Hugoniot  jump
conditions for a  shock with isothermal Mach  number $M_s$ propagating
into  the medium  surrounding the  cloud.  We  considered a  number of
different cases with different external  pressures and sound speeds as
summarised in Table \ref{table1}.

The only dimensionless parameters are  the incident shock Mach number,
$M_s$,  the ratio  of the  initial  external pressure,  $P_e$, to  the
maximum pressure  of a stable  Bonnor-Ebert sphere, $P_{max}$  and the
ratio of the sound speeds, $c_c$  and $c_e$. In the case considered by
\cite{Li:2014}, the cloud had a  mass of $1$ M$_{\sun}$, a temperature
of $10$ K, and a radius of  $0.058$ pc in an external medium at $1000$
K. This gives  the same ratio of  sound speeds as in our  State 1 (see
table see Table \ref{table1}), but their initial external pressure was
$0.914 P_{max}$, so  that their cloud was closer to  collapse than any
of our  cases.  It is therefore  not surprising that they  find that a
shock with  a Mach number  of $1.5$  induces collapse. One  could also
apply our results to other cases,  such as Bok globules with masses in
the range $10$  -- $100$ M$_{\sun}$, which are known  to be associated
with young stars \cite{Yun:1990}.

It is useful to define an incident shock crossing time by

\begin{equation}
t_c = \frac{R}{ c_e M_s},
\label{tcross}
\end{equation}

\noindent
where  $R$ is  the  radius  of the  equilibrium  cloud,  which is  the
timescale   on   which   the    flow   outside   the   cloud   becomes
quasi-steady. The other relevant timescale  is an estimate of the time
it takes for  the transmitted shock in the cloud  to reach the centre,
which \cite{Klein:1994} call the ``cloud crushing time''. They use

\begin{equation}
t_{cc} = t_c \frac{\rho_{cloud}}{\rho_e},
\label{tcrush1}
\end{equation}

\noindent
where    $\rho_{cloud}$   is    the   (uniform)    density   of    the
cloud. \cite{Li:2014}  set $\rho_{cloud}  = \rho_c$ where  $\rho_c$ is
the central density. We shall see later that this is not always a good
estimate of the time at which  the maximum density in the cloud begins
to increase.

Figure~\ref{fig1} shows the  density and velocity vectors in  the $z =
0$ plane  at different times for  State 1, $M_s =  2.2$ simulation. In
the  top left  hand panel  ($t =  3.359 t_c$)  the incident  shock has
passed the  cloud, formed a  quasi-steady bow  shock and is  driving a
transmitted shock  into the  cloud.  In  the middle  left panel  ($t =
11.292 t_c$) the transmitted shock has reached the centre of the cloud
in the cloud and created a high density region. The middle right panel
is at the time at which the  density is maximum ($t = 14.662 t_c$) and
in the bottom left panel ($t = 17.426 t_c$) the cloud is re-expanding.
The  bottom right  panel ($t  = 23.138  t_c$) shows  a blow-up  of the
expanding cloud with velocity arrows in  the rest frame of the densest
region.

The convergence  of the transmitted  shock at  the centre is  shown in
more  detail in  figure~\ref{fig2}.  Here  we can  see that  a jet  is
produced   in   much   the   same   way  as   in   a   shaped   charge
\citep{Birkhoff}. This jet interacts with  other parts of the cloud to
produce  a secondary  high density  region (bottom  left panel).   The
maximum  density  is   much  larger  than  that   behind  the  initial
transmitted shock: it is clear that flow convergence has a significant
effect There are some similarities with the magnetic case described in
\cite{Vaidya:2013p12979}, but  the latter  is more complicated  due to
the dynamic effects  of the magnetic field. However,  despite the high
density, there  is no collapse in  this case because the  high density
region is too small to be gravitationally unstable.

Figure~\ref{fig3} shows  what happens for State  1 in the $M_s  = 2.3$
case, which  does collapse.  The flow  evolves in  much the  same way,
except  that  the  high  density region  collapses in this case.

Table ~\ref{table1} shows that the Mach number of a shock that induces
collapse does  not decrease by much  when the external sound  speed is
reduced to $4$ (State 2) and in  fact the flow is very similar to that
for  State 1.   This is  not too  surprising since  in both  cases the
density contrast is large enough for  the evolution of the flow around
the  cloud to  be quasi-steady.   The  values of  the maximum  density
obtained for  different simulations are listed  in Table~\ref{table1}.
One  can see  from Figure~\ref{fig7}  that the  cloud bounces  even in
those cases in which it subsequently undergoes collapse.

It would obviously be useful to have  a rule of thumb to determine the
strength  of  the  shock  required for  collapse.   Consider  a  plane
isothermal shock with  speed $s$ in a medium at  rest with sound speed
$c$, density $\rho_e$ and pressure $P_e = c^2 \rho_e$. The density and
velocity behind the shock are

\begin{equation}
\rho_1 = \rho_e M^2,
\label{shock1}
\end{equation}

\noindent
and

\begin{equation}
v_1 = s \left( {1 - \frac{1}{M^2} } \right).
\label{shock2}
\end{equation}

The pressure is therefore

\begin{equation}
P_1 =  M^2 c^2 \rho_e = M^2 P_e.
\label{shock3}
\end{equation}

The most obvious estimate of the  critical Mach number for collapse is
to set $P_1  = P_{max} = 1.4$, the maximum  external pressure that the
cloud can  support, which gives the  Mach number $M_1$ shown  in Table
~\ref{table1}. One  can see that this  works quite well for  the cases
with the larger  initial external pressure (States 4 and  5) but is an
overestimate for the  lower initial pressues (States 1, 2  and 3). One
might suppose  that this is  because shock convergence  induces larger
pressures than that behind the incident shock.

However, as  we have already  pointed out,  the evolution of  the flow
around  the  cloud  to  approximately quasi-steady,  so  the  relevant
pressure should really  be that in a steady flow  around the cloud. In
such a  flow, the maximum pressure  is that behind a  stationary shock
whose upstream state is that behind  the incident shock, i.e $v = v_1$
and $\rho = \rho_1$. The post-shock density and pressure are therefore

\begin{equation}
\rho_2 = \rho_1 \left(\frac{v_1}{c} \right)^2 = \rho_e M^4 \left( {1 -
  \frac{1}{M^2} } \right)^2.
\label{shock4}
\end{equation}

\begin{equation}
P_2 = \frac{\rho_2}{\rho_e} P_e =  M^4 \left( {1 -
  \frac{1}{M^2} } \right)^2 P_e.
\label{shock5}
\end{equation}

Setting $P_2  = P_{max} =  1.4$ gives the  Mach number $M_2$  shown in
Table  ~\ref{table1}.   This works  very  well  for  State 3,  and  is
somewhat better than $M_1$ for States 1 and 2.  Note that $M_2$ is not
defined for States 4 and 5 since it would imply a subsonic flow behind
the  incident shock.   In  fact  the State  4  and  5 simulations  are
somewhat  dubious  since  the  flow is  subsonic  behind  the  weakest
incident shock that  induces collapse, which is  incompatible with our
imposition of the post-shock state at the right x boundary.

\begin{figure}
\includegraphics*[viewport=1050 291 1920 1080,width=1.0\columnwidth]{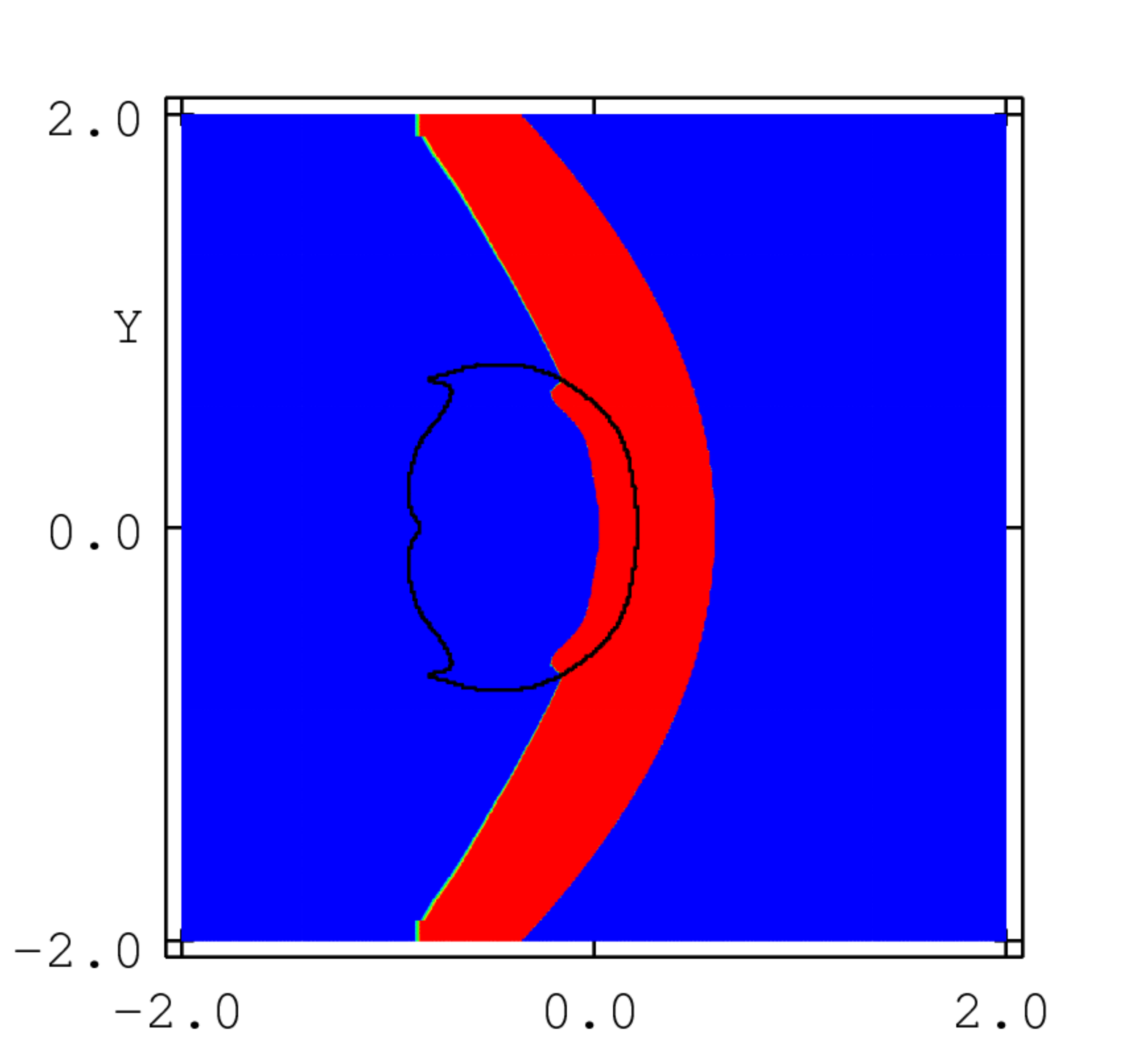}
\caption{Red  shows the  region  where the  pressure  is greater  than
  $P_{max}$  for the  State 1  $M_s =  2.3$ calculation  at $t  = 5.49
  t_c$. The boundary of the cloud is the black contour.}
\label{fig4}
\end{figure}

Nevertheless,  $M_2$  significantly  underestimates  the  Mach  number
required for collapse for States 1  and 2. Figure ~\ref{fig4} tells us
that for State 1, only a fairly small part of the surface of the cloud
experiences a pressure greater than $P_{max}$ and the same is true for
State 2.  Clearly shock convergence helps,  but it does not  produce a
large enough high density region for collapse unless the incident Mach
number is somewhat higher than $M_2$.

\begin{figure}
\includegraphics*[viewport=1050 265 1920 1080, 
width=1.0\columnwidth]{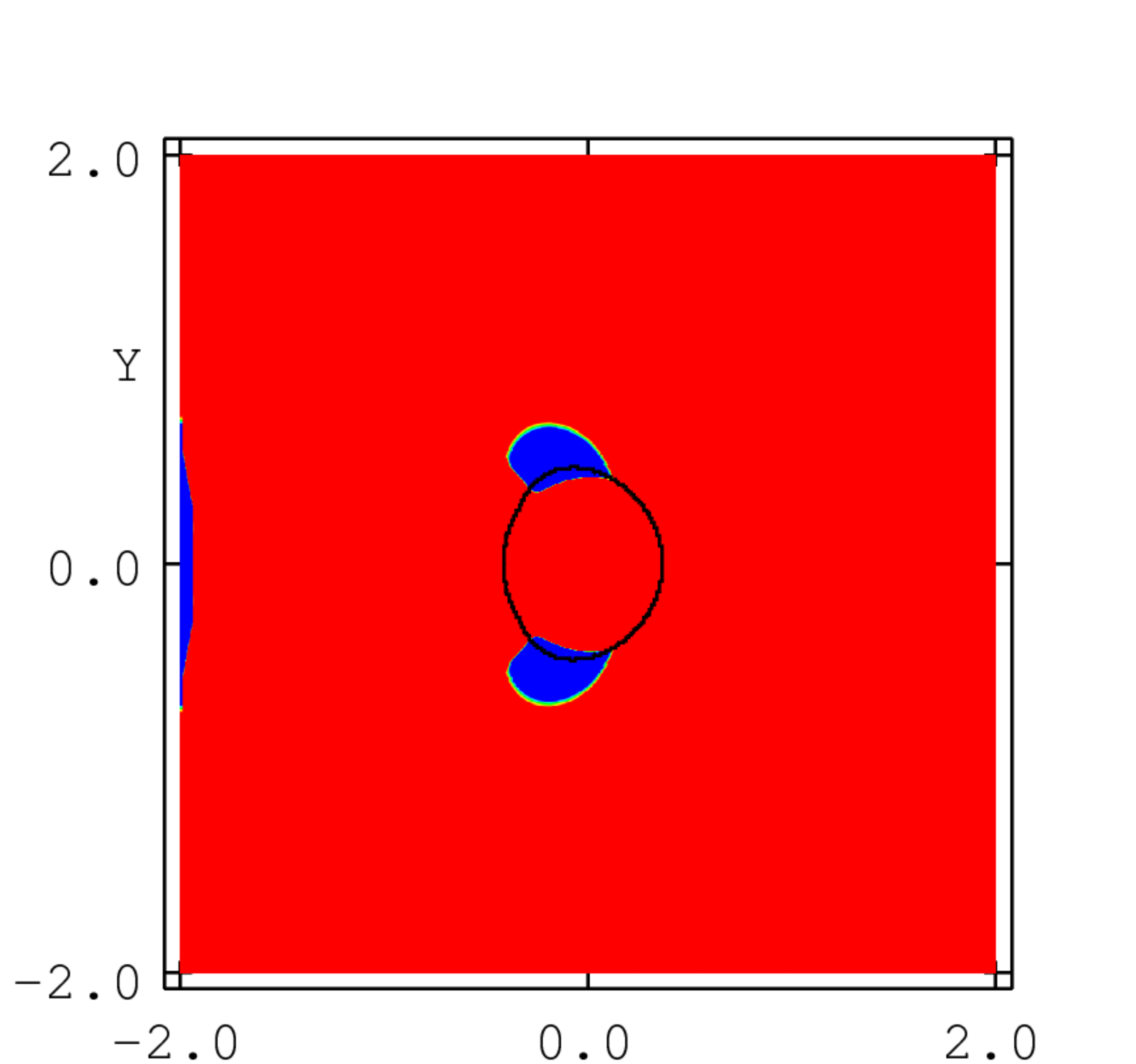}
\caption{Red  shows the  region  where the  pressure  is greater  than
  $P_{max}$  for the  State 5  $M_s =  1.3$ calculation  at $t  = 4.82
  t_c$. The boundary of the cloud is the black contour.}
\label{fig5}
\end{figure}

One  can see  from Figure  ~\ref{fig5} that  almost the  whole of  the
surface of the cloud experiences  a pressure greater than $P_{max}$ in
the subsonic  flow behind the incident  shock for State 5  and this is
also true for State 4.  This is why $M_1$ is a good estimate for these
cases.  However, as  we have already pointed out  the calculations are
not entirely trustworthy for these cases and even if they were, clouds
so close to collapse are not of much interest.

\begin{figure}
\includegraphics*[viewport=0 0 234 234,
width=1.0\columnwidth]{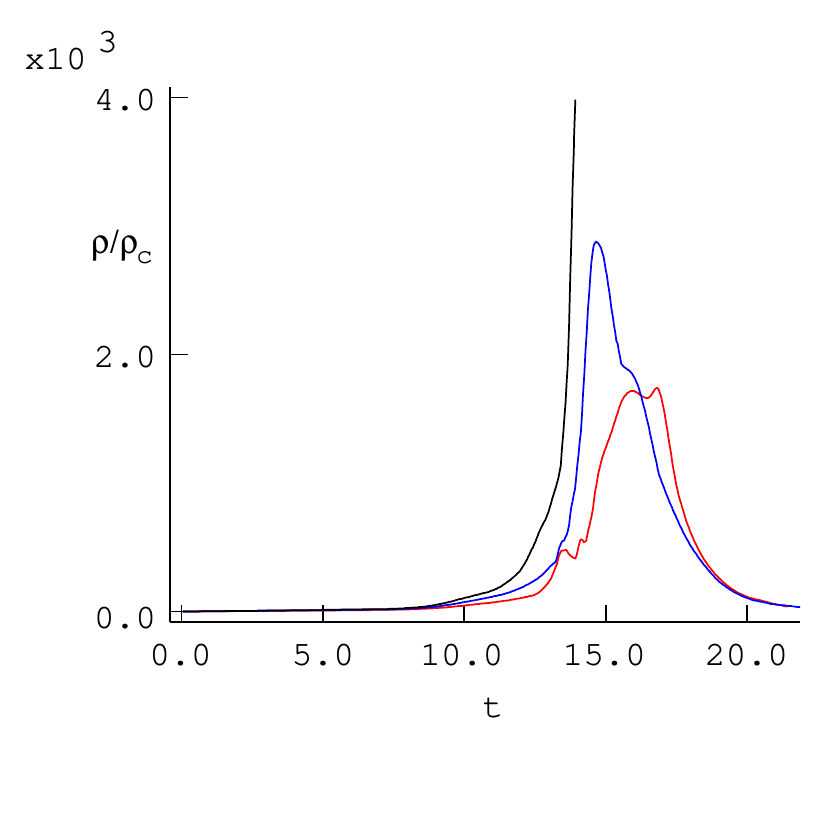}
\caption{Ratio of the maximum  density to the initial central density,
 $\rho_c$, for State 1 as functions  of time. $M_s = 2.1$ -- red; $M_s
 = 2.2$ -- blue; $M_s = 2.3$ -- black (see Table~\ref{table1}).}
\label{fig6}
\end{figure}

\begin{figure}
\includegraphics*[viewport=0 0 234 234,
width=1.0\columnwidth]{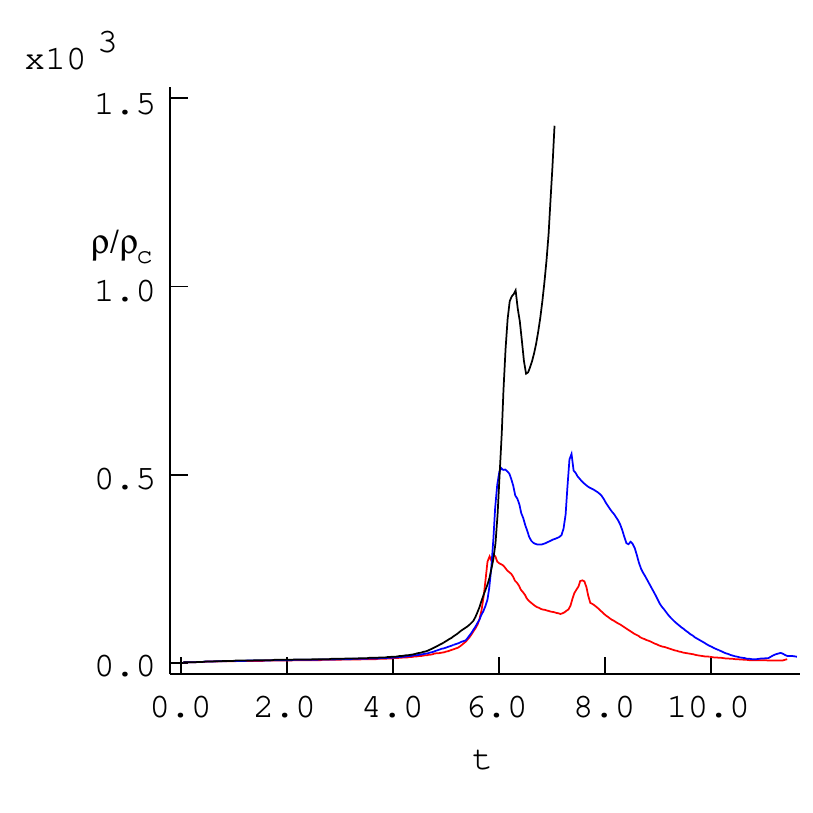}
\caption{Ratio of the maximum  density to the initial central density,
 $\rho_c$, for State 2 as functions of time. $M_s = 2.0$ -- red;
 $M_s  =  2.1$  --  blue;  $M_s  =  2.2$  --  black  (see
 Table~\ref{table1}).}
\label{fig7}
\end{figure}

Note that neither $M_1$ nor $M_2$  depend on the initial density ratio
between  the cloud  and its  surroundings, $c_c^2/c_e^2$,  since $M_1$
simply depends on the pressure behind  the incident shock and $M_2$ on
the pressure behind  a quasi-steady bow shock.  The  critical value of
$M_s$ for  collapse is also  insensitive to  the density ratio:  it is
very nearly the same for States 1  and 2.  This is because the density
ratio  is  large enough  for  the  ``cloud  crushing time''  given  by
equation  (\ref{tcrush1}) to  be significantly  larger than  the shock
crossing time, given by equation  (\ref{tcross}). The flow outside the
cloud  is  therefore  approximately   quasi-steady  by  the  time  the
transmitted shock  reaches the centre of  the cloud.  We can  see that
this is true from Figures \ref{fig6} and \ref{fig7}: the cloud density
does not  begin to increase  signicantly until  $t \simeq 10  t_c$ for
State  and 1  $t \simeq  5 t_c$  for State  2. Note  that the  ``cloud
crushing time'' defined by equation (\ref{tcrush1}) gives a reasonable
estimate of this time:  $t_{cc} = 13.6 t_c$ for State  1 and $t_{cc} =
5.4 t_c$ for State  2. However, the estimate is not  so good for State
5, for which $t_{cc} = 8.1  t_c$, while the density begins to increase
rapidly at $t \simeq 2.5 t_c$.   This is largely because this cloud is
more centrally condensed,  so that the speed of  the transmitted shock
in the outer parts  of the cloud is larger than  that used to estimate
$t_{cc}.$

\section{Hydrodynamic Instabilities}
\label{instab}

There has recently been considerable  work on the possibility that the
short  lived  radio  isotopes  found in  meteorites  could  have  been
injected into the solar nebular by instabilities during an interaction
with  a  supernova  shell (e.g.  \citealt{Boss:2012p12244,  Li:2014}).
Although it is not the main focus of the paper, it is worth looking at
any instabilities that  might occur.

\begin{figure}
\includegraphics[width=1.\columnwidth]{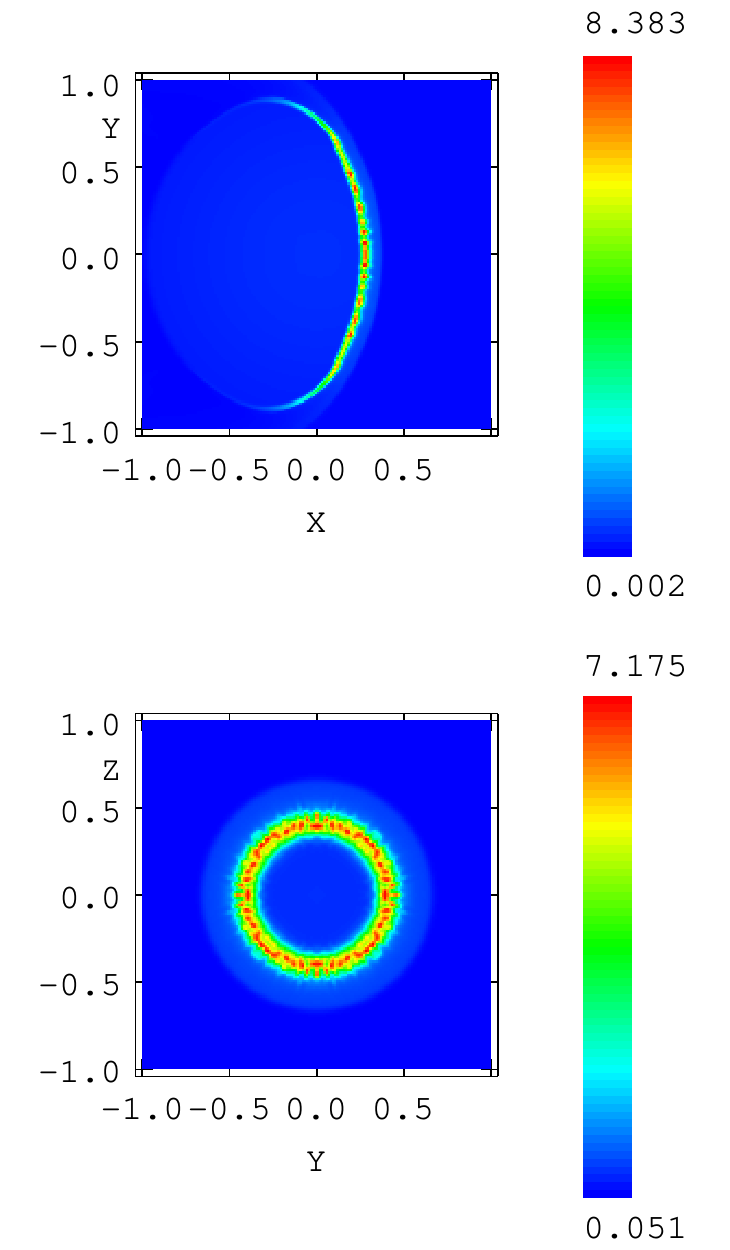}
\caption{Density  in the  $z =  0$  plane (top)  and $x  = 0.2$  plane
  (bottom) for  the State  1,  $M_s =  5.0$ simulation  at  $t =  2.398
  t_c$.}
\label{fig8}
\end{figure}

\begin{figure}
\includegraphics[width=1.\columnwidth]{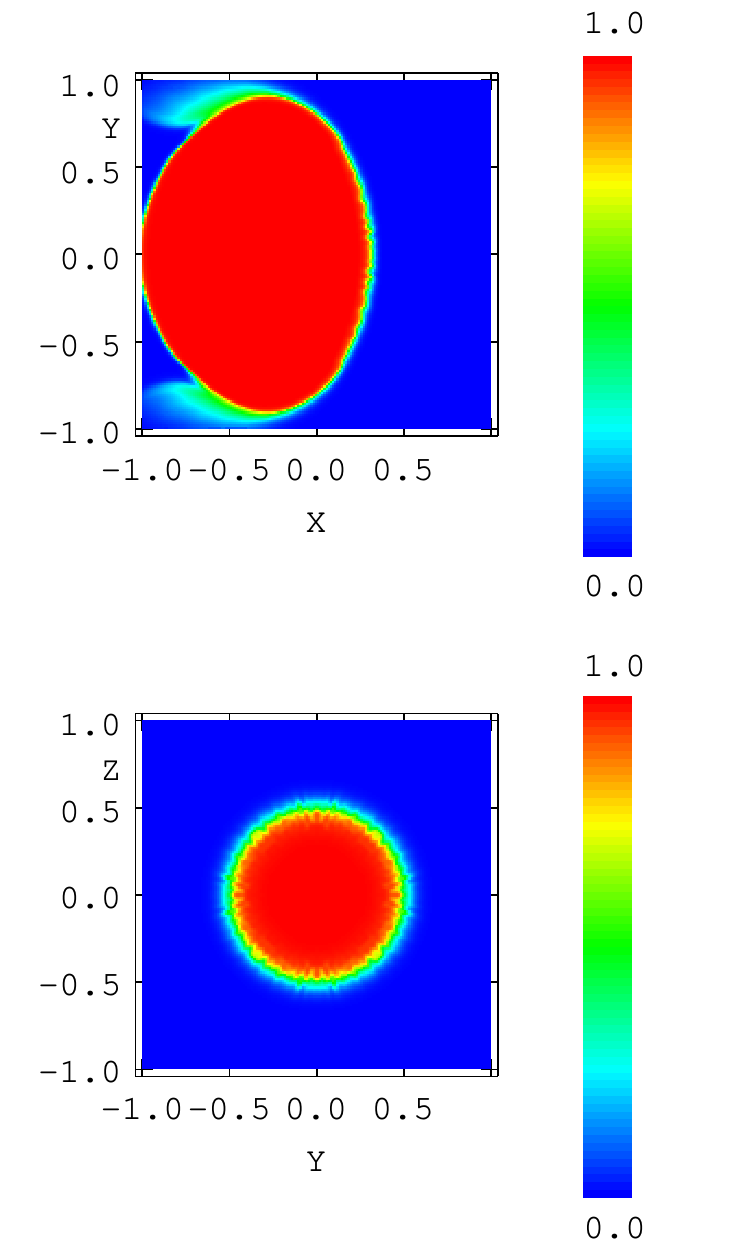}
\caption{Cloud scalar in the  $z = 0$ plane (top) and  $x = 0.2$ plane
  (bottom)  for the  State 1,  $M_s =  5.0$ simulation  at $t  = 2.398
  t_c$.}
\label{fig9}
\end{figure}

One  might  expect  self-gravity  to  have  a  stabilising  effect  on
instabilities at the cloud surface since it is Rayleigh-Taylor stable.
The  simulations described  in  the previous  section  do indeed  show
little  evidence  of   either  Kelvin-Helmholtz  or  Richtmyer-Meshkov
instabilities   at  the   cloud   surface,  but   there   is  a   weak
non-axisymmetric Kelvin-Helmholtz instability in the wake.  However, a
stronger incident shock does induce Richtmyer-Meshkov instabilities on
the front surface  of cloud.  Figure ~\ref{fig8} shows  the density in
the $z  = 0$  and $x  = 0.2$  planes for a  State 1,  $M_{\rm s}  = 5$
simulation at a  time at which the incident shock  has just passed the
cloud.   A non-axisymmetric  Richtmyer-Meshkov instability  is clearly
present and has not been suppressed by self-gravity.  One can see from
Figure \ref{fig9}, which shows the  cloud scalar, that the instability
does lead to some mixing behind  the shock in the cloud. However, this
shock simply shreds the  cloud without causing gravitational collapse.
This is consistent with the  results in \cite{Boss:2012p12244}: if the
shock is too  weak there is very little instability,  whereas if it is
too strong it destroys the cloud.

\section{Conclusions}
\label{conclusion}

We have  found that the  minimum Mach number,  $M_s$, of a  shock that
induces collapse in a previously  stable self-gravitating cloud is not
sensitive  to the  initial density  ratio  between the  cloud and  its
surroundings, as long as it  is sufficiently large.  The mininum value
of  $M_s$ lies  between $M_2$  and  $M_1$ where  for $M_s  = M_1$  the
pressure behind the  incident shock is equal to  the maximum pressure,
$P_{max}$, that cloud can sustain, whereas for $M_s = M_2$ the maximum
pressure behind  the bow  shock is  equal $P_{max}$.   A shock  with a
smaller value of  $M_s$ will compress a cloud until  a peak density is
reached, after which  the cloud expands, but gravity  prevents it from
being disrupted.  Such  an expansion phase can occur even  if $M_s$ is
large enough for collapse, but in such cases gravity eventually drives
collapse.

As  in  our  simulations   of  shock  interactions  with  magnetically
subcritical clouds \cite{Vaidya:2013p12979}, the simulations described
above show that shock focussing is  responsible for the large value of
the peak density that is reached even in clouds that do not collapse.
 
\section*{Acknowledgements}

This  work  was supported  by  the  Science \&  Technology  Facilities
Council  (Research Grant  ST/I001557/1).   The  calculations for  this
paper were performed  on the DiRAC 1 Facility jointly  funded by STFC,
the  Large  Facilities Capital  Fund  of  BIS  and the  University  of
Leeds.  This  facility is  hosted  and  enabled  through the  ARC  HPC
resources  and support  team  at  the University  of  Leeds (A.  Real,
M.Dixon, M. Wallis,  M.  Callaghan \& J. Leng), to  whom we extend our
grateful thanks. We thank an anonymous referee for helpful comments.

\bibliographystyle{mn2e}


\label{lastpage}

\end{document}